\documentclass[12pt]{JHEP}
\usepackage{graphics}
\usepackage{cite}
\newcommand{\beq}{\begin{equation}}
\newcommand{\eeq}{\end{equation}}
\newcommand{\bea}{\begin{eqnarray}}
\newcommand{\eea}{\end{eqnarray}}

\renewcommand{\d}{\delta}

\renewcommand{\b}{\beta}

\newcommand{\m}{\mu}

\newcommand{\s}{\sigma}

\newcommand{\oh}{\frac{1}{2}}

\newcommand{\rf}[1]{(\ref{#1})}
\newcommand{\ra}{\rightarrow}

\title{P-Vortices, Gauge Copies, and Lattice Size}

\author{Roman Bertle \\ Inst. f\"ur Kernphysik, 
Technische Universit\"at Wien, A-1040 Vienna, Austria \\
E-mail: \email{bertle@kph.tuwien.ac.at}}

\author{Manfried Faber \\ Inst. f\"ur Kernphysik, 
Technische Universit\"at Wien, A-1040 Vienna, Austria \\
E-mail: \email{faber@kph.tuwien.ac.at}}

\author{Jeff Greensite \\ The Niels Bohr Institute,
Blegdamsvej 17, DK-2100 Copenhagen \O, Denmark \\
E-mail: \email{greensit@alf.nbi.dk} }

\author{{\v S}tefan Olejn\'{\i}k \\ Institute of Physics, Slovak Academy 
of Sciences, SK-842 28 Bratislava, Slovakia \\
E-mail: \email{fyziolej@savba.sk}}

\abstract{We study the dependence of the center-projected string
tension on both the lattice size, and the number of gauge copies used for
maximal center gauge fixing. We show that a recent finding of
Bornyakov, Komarov, Polikarpov, and Veselov (hep-lat/0002017),
indicating a substantial breakdown of center dominance in maximal
center gauge, is only obtained for rather small
lattice volumes, and is not found in numerical simulations on larger
lattices.  It is shown that center-projected Creutz ratios approach
the full asymptotic string tension as lattice size increases, and that
the P-vortex density is consistent, at moderately weak couplings, with
2-loop scaling behavior.}


\preprint

\begin{document}

\section{Introduction}

   In the last few years there has been renewed interest, within the
lattice gauge theory community, in the center vortex theory of
confinement \cite{tH}.  The revival of this old idea is due to a
number of numerical studies, which all indicate that center vortices
are ubiquitous in the QCD vacuum and give rise to the linear
confining potential \cite{indirect,Zako,Jan98,mog,dFE,dFE1,Lang1,
Tubby,Tubby1,Lang2,bertle,ITEP,ET,Hart}.  Recently, however, a paper by
Bornyakov, Komarov, Polikarpov, and Veselov (BKPV) \cite{Borny}
appeared which questions the validity of a procedure, known as center
projection in maximal center gauge, which was essential in many of
these studies.  The claim is that when large numbers of random gauge copies
are used in fixing the gauge (in an effort to minimize the Gribov copy
problem), the center-projected string tension underestimates the
full string tension by as much as $30\%$.  Our purpose in the present
article is to show that, while BKPV have certainly raised an important
issue, their actual data was strongly affected by the rather small
lattice volumes used in the numerical simulations.  We will show that the
conclusions drawn from this data are not sustained by simulations on
larger lattices, which generally agree with our previously reported
results \cite{Jan98}. 

   In the direct version of maximal center gauge \cite{Jan98}, the 
procedure is to maximize
\beq
       R = \sum_\m \sum_x \mbox{Tr}_A[U_\m(x)]
\eeq   
by an iterative over-relaxation procedure, where
$\mbox{Tr}_A[U]$ is the trace of $U$ in the adjoint representation.
Let $R_n$ denote the value of $R$ after $n$ over-relaxation sweeps.
When $R_n$ is judged to have converged, e.g.\ according to a criterion 
of the form
\beq
       {R_n - R_{n-50} \over R_n} < \d
\label{delta}
\eeq
then the link variables $U_\m(x)$ are projected onto the nearest
center element $Z$.  In SU(2) lattice gauge theory, the projection is
simply
\beq
       Z_\m(x) = {\rm signTr}[U_\m(x)]
\eeq
Center-projected Wilson loops, Creutz ratios, etc.\ are observables
computed from the center-projected link variables.  It was shown in a
number of studies (see, in particular, ref.\ \cite{Jan98}), that 
\begin{itemize}
\item thin vortex excitations of the projected lattice, known as
``P-vortices'', are located roughly in the middle of thick center
vortices on the unprojected lattice;
\item projected Creutz ratios $\chi_{cp}(I,I)$ are close to the asymptotic
string tension on the unprojected lattice (``center dominance'');
\item the density of P-vortices, from $\b=2.3$ onward, scales according
to asymptotic freedom;
\item removing center vortices (located via the projected lattice)
from unprojected lattices also removes confinement and chiral symmetry
breaking, and brings the topological charge on the lattice to zero \cite{dFE}. 
\end{itemize}

   Because these numerical results have such strong implications for
the QCD confinement mechanism, it is important that the center
projection procedure be examined critically.  The first obvious
question $-$ why should this procedure work at all? $-$ was addressed
in ref.\ \cite{vf}.  There it was shown that in the absence of Gribov
copies (i.e.\ if the gauge can be fixed to a global maximum of $R$),
then center projection in maximal center gauge will always locate a
thin vortex inserted anywhere on the lattice.  This was dubbed the
``vortex-finding property'' of maximal center gauge, and is certainly
a necessary condition for its success.  However, the vortices found in
the QCD vacuum are not thin vortices, but are necessarily of finite
thickness in physical units.  Moreover, maximal center gauge is
plagued with Gribov copies, since the over-relaxation scheme converges
only to a local maximum of $R$, which will be slightly different for
every gauge copy of a given lattice configuration.

   In view of the Gribov copy problem, and
the finite thickness of vortices, one must rely on empirical checks of
vortex-finding via center projection.  In this article we will study
the sensitivity of center-projected Creutz ratios and the P-vortex density
with respect to: (i) the number $N_{copy}$ of random gauge copies used for 
maximal center gauge fixing; (ii) the lattice size; and
(iii) the convergence parameter $\d$ in eq.\ \rf{delta}.
We will here only discuss center projection in direct maximal center gauge,
since this is the case treated in ref.\ \cite{Borny}, and it is also
the center gauge with which we have the most experience.  It should be
noted, however, that alternatives to maximal center gauge do exist; 
these include Laplacian center gauge \cite{dFE1} (which is free of
the Gribov copy problem), as well as two other
recent proposals \cite{Zub,Lang3}.

\section{\boldmath $N_{copy}$ and Lattice Size Dependence}

   We begin with the $N_{copy}$ dependence, whose importance was
recently emphasized by BKPV \cite{Borny}.  As noted above, when the
over-relaxation gauge-fixing procedure is applied to different gauge
copies of a given lattice configuration, slightly different values of
$R$ are obtained. One way to minimize the gauge copy dependence is to
carry out the over-relaxation procedure on a number $N_{copy}$ of
random gauge copies, perform center projection on the copy with the
largest value of $R$, and evaluate observables.  Data obtained in this
way, over a range of $N_{copy}$ values, can then be extrapolated to
the $N_{copy}\ra \infty$ limit.\footnote{P-vortex positions extracted
from different Gribov copies are closely correlated, as one would
expect if they locate physical objects \cite{Jan98}. It is possible,
however, by choosing a very special starting configuration with links
gauge-fixed to have mainly positive trace, to destroy the
vortex-finding property of center projection even for thin vortices
\cite{KT,vf}. This is one aspect of the Gribov copy problem in maximal
center gauge. Since the vortex-finding property is essential to center
projection, it is evident that any unusual variation of the iterative
gauge-fixing procedure, which destroys this property, must be
avoided.}  In the original simulations of ref.\ \cite{Jan98}, only
three gauge copies were used, and there was no attempt to extrapolate
to $N_{copy}\ra \infty$. Projected Creutz ratios $\chi_{cp}(I,I)$ were
found to be close to the asymptotic string tensions reported by Bali
et al.\ \cite{Bali}, for all $I\ge 2$.  BKPV, however, calculate
Creutz ratios in the range $N_{copy} \in [1,20]$, and extrapolate to
$N_{copy} \ra \infty$ by fitting their data to the functional form
\beq
   \chi^{N_{copy}}_{cp}(I,I) = \chi_{cp}(I,I) + 
       {c(I,I) \over \sqrt{N_{copy}} } 
\label{fit}
\eeq
The result reported by BKPV is that projected string tensions, at
$\b=2.4,2.5$, underestimate the full string tension by about $20\%$
at $N_{copy}=20$, and by as much as $30\%$ in the extrapolation to
$N_{copy}\ra \infty$.  This result suggests that the center
dominance previously reported in maximal center gauge was 
a numerical accident, a result of using too few gauge copies.

   However, apart from the number of gauge copies, there is one other
notable difference between the BKPV simulations and previous work.
This is the matter of lattice size.  Data reported in ref.\
\cite{Jan98} was obtained at $\b=2.3$ and $\b=2.4$ on $16^4$
lattices, and at $\b=2.5$ on a $22^4$ lattice.  BKPV, on the other
hand, used only $12^4$ lattices at $\b=2.3$ and $\b=2.4$, and lattice
size $16^4$ at $\b=2.5$.  This raises the question of whether the BKPV
results, obtained on the smaller lattices, were seriously contaminated
by finite-size effects.  To find out, we have repeated the
center-projection calculation at $\b=2.3$ and $\b=2.5$ on a variety of
lattice sizes, for $N_{copy}\in [1,20]$.  For the convergence parameter
in eq.\ \rf{delta}, we have used $\d=2\times 10^{-7}$.

\FIGURE[h]{
\centerline{\scalebox{0.9}{\includegraphics{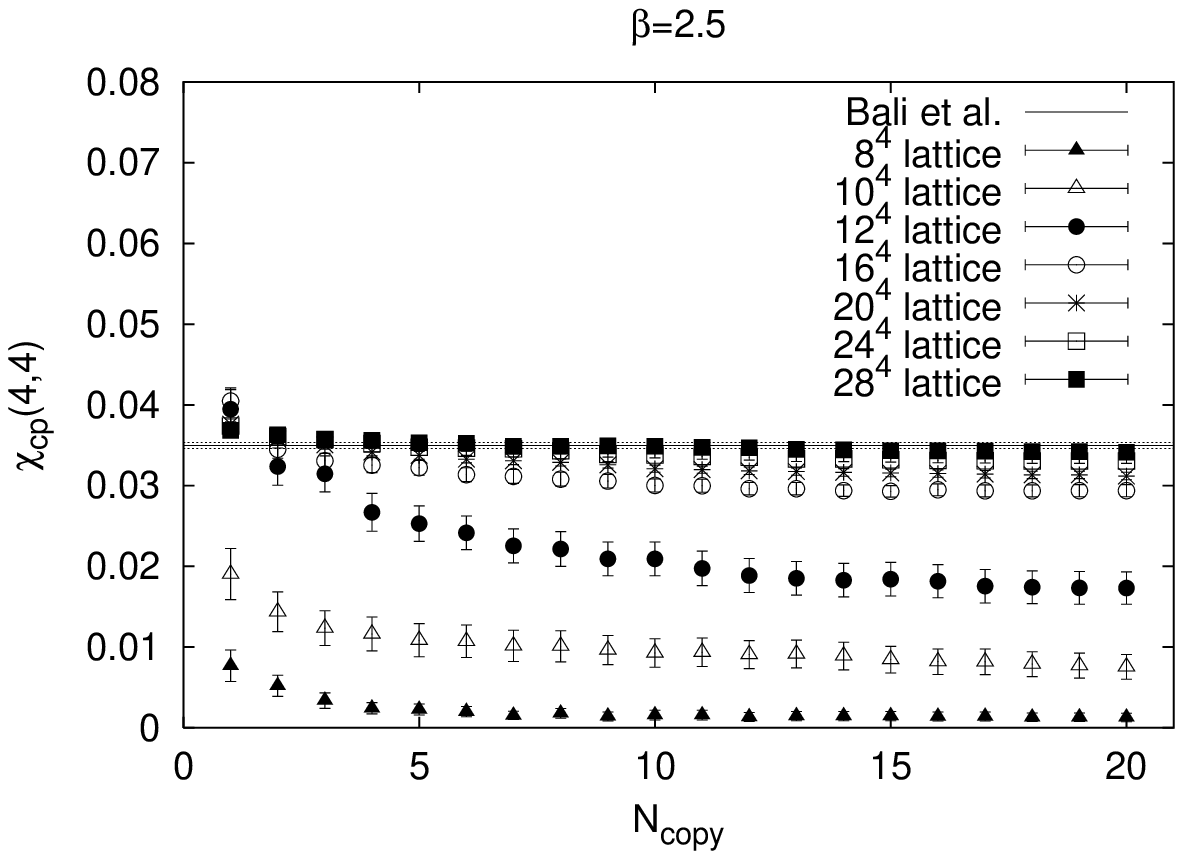}}}
\caption{Center projected Creutz ratios $\chi_{cp}^{N_{copy}}(4,4)$ 
vs.\ $N_{copy}$ at various lattice volumes, for $\b=2.5$.  The 
solid line in this figure, and in Figs.\ \ref{more23}-\ref{svh}, 
indicates the asymptotic 
string tension extracted 
by standard methods on unprojected lattices, reported by Bali
et al.\ \cite{Bali}.  Dashed lines indicate the errorbar in this
asymptotic string tension.}
\label{fig1}
}    

   In Fig.\ \ref{fig1} we display results for the Creutz ratio 
$\chi_{cp}^{N_{copy}}(4,4)$ vs.\ $N_{copy}$ at $\b=2.5$, 
for lattice sizes ranging from $8^4$ to $28^4$.  Two features of this
data are immediately apparent.  First, there is indeed a slow
downward trend in the Creutz ratio as $N_{copy}$ increases, as noted
by BKPV, but this
effect is much more pronounced on smaller lattices than on larger lattices.
Second, although Creutz ratios on the smaller lattices grossly underestimate
the full string tension, the data appears to steadily increase towards the 
full asymptotic string tension, reported by Bali et al.\ \cite{Bali},
as the lattice size increases.  These trends in the data are by no means
unique to the particular Creutz ratio $\chi_{cp}(4,4)$ at $\beta=2.5$, but
are typical of all of our results.  For completeness we display, in 
Figs.\ \ref{more23} and \ref{more25}, some other projected Creutz ratios 
$\chi_{cp}^{N_{copy}}(I,I)$ for $I$ in the range $I=2-5$, at couplings
$\b=2.3$ and $\b=2.5$.  In Figs.\ \ref{more23}-\ref{svh}, solid (dashed)
lines indicate the value (errorbar) of the asymptotic string tension
on the unprojected lattice, reported in ref.\ \cite{Bali}.

\FIGURE[h]{
\begin{tabular}{cc}
\scalebox{0.6}{\includegraphics{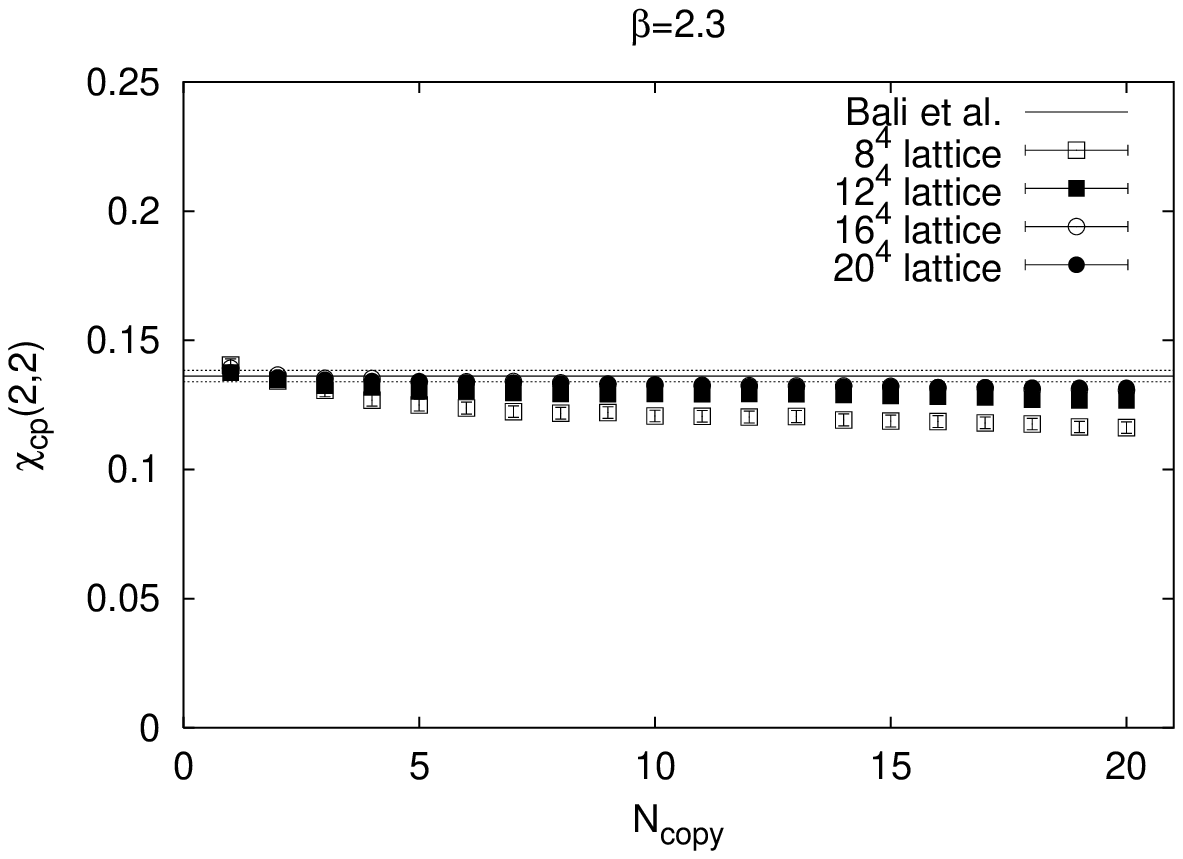}} &
\scalebox{0.6}{\includegraphics{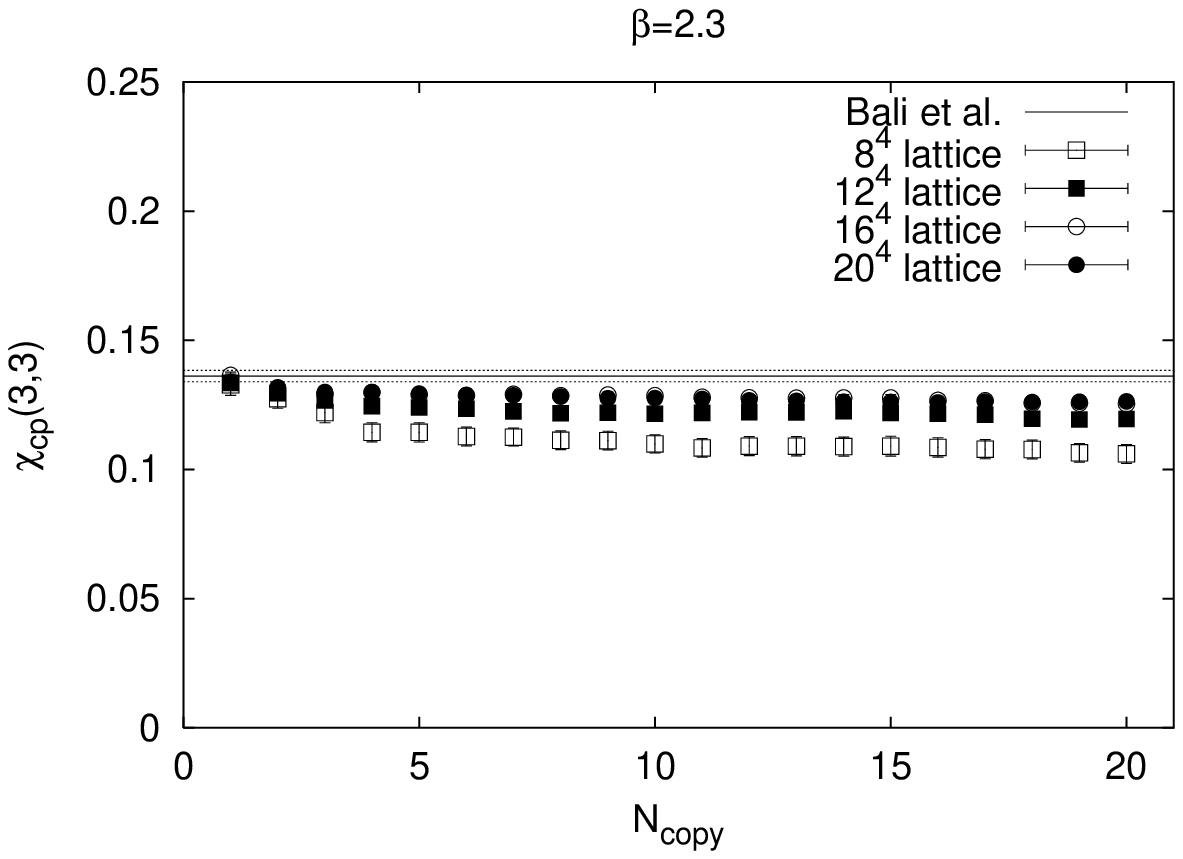}} \\
\scalebox{0.6}{\includegraphics{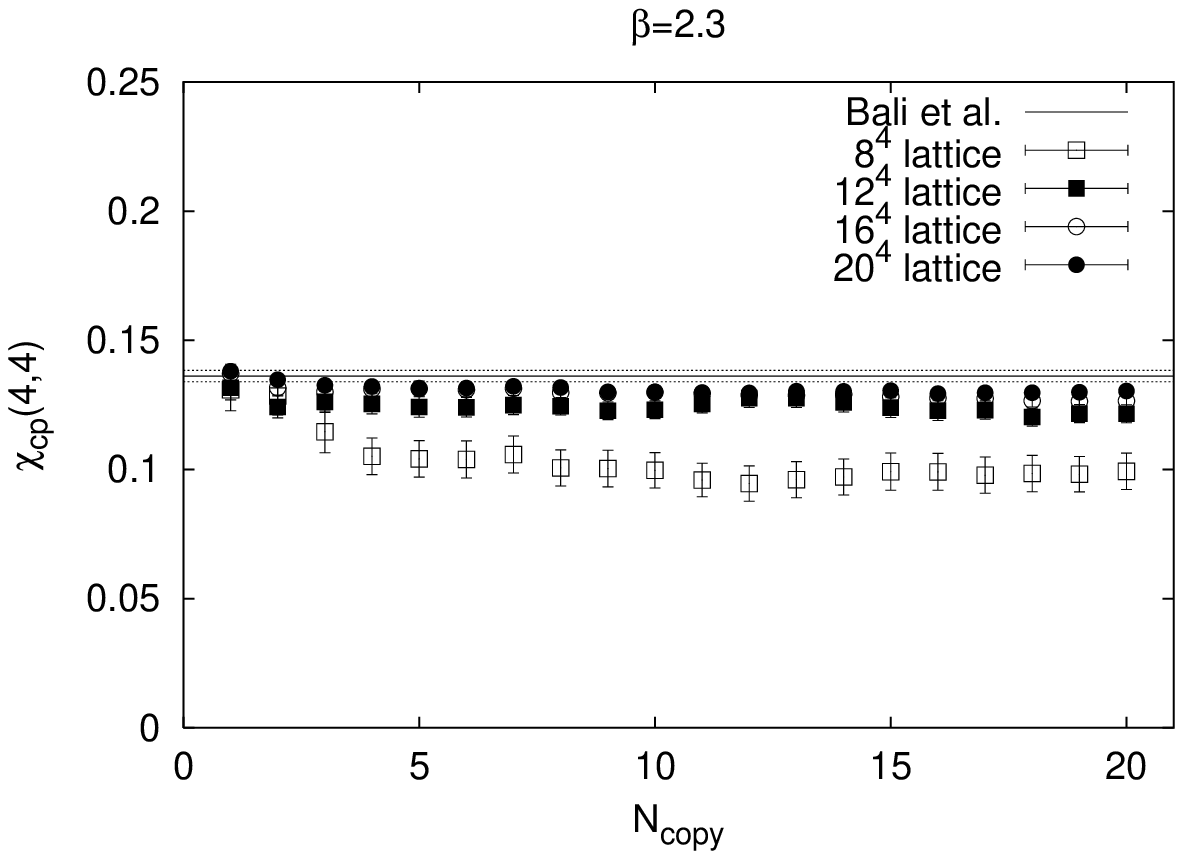}} &
\scalebox{0.6}{\includegraphics{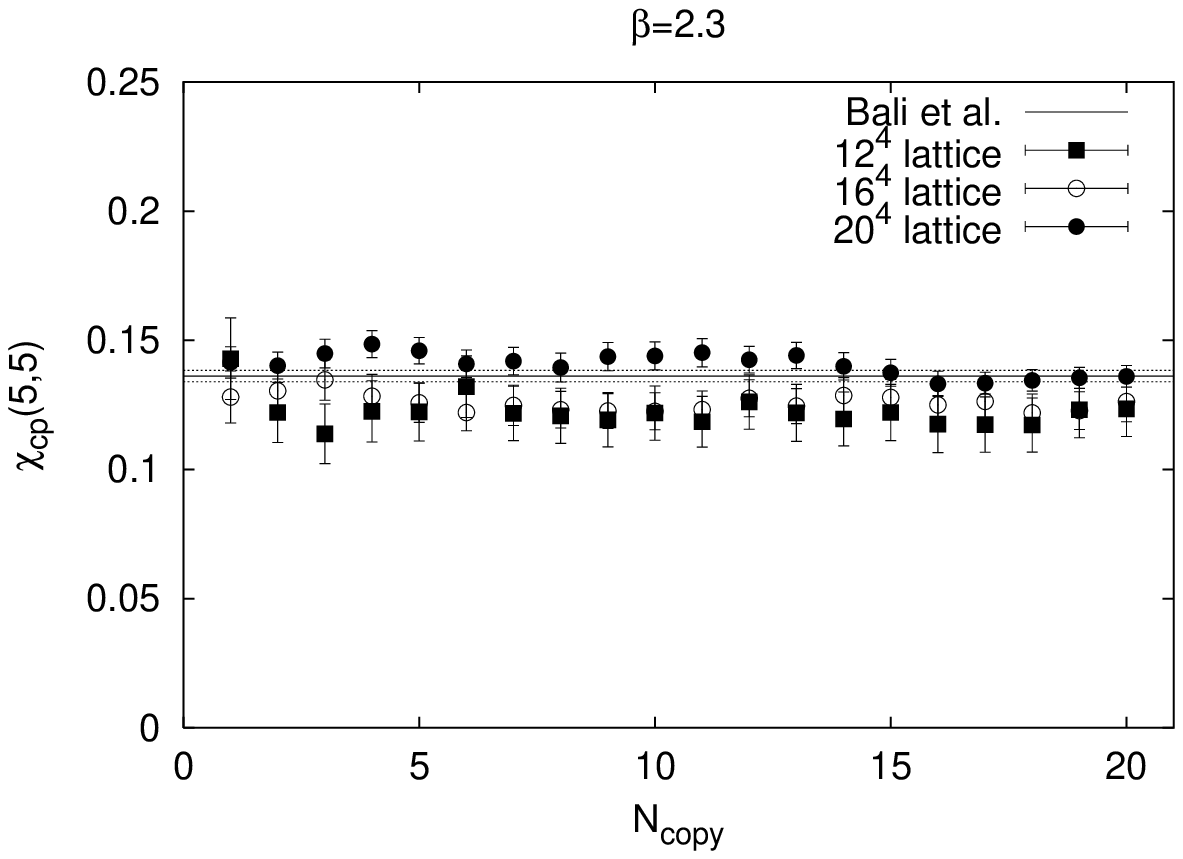}} \\
\end{tabular}
\caption{Center projected Creutz ratios $\chi_{cp}^{N_{copy}}(I,I)$ 
vs.\ $N_{copy}$ and lattice volume, at $\b=2.3$ and $I=2-5$.}
\label{more23}
}   

\FIGURE[h]{
\begin{tabular}{cc}
\scalebox{0.6}{\includegraphics{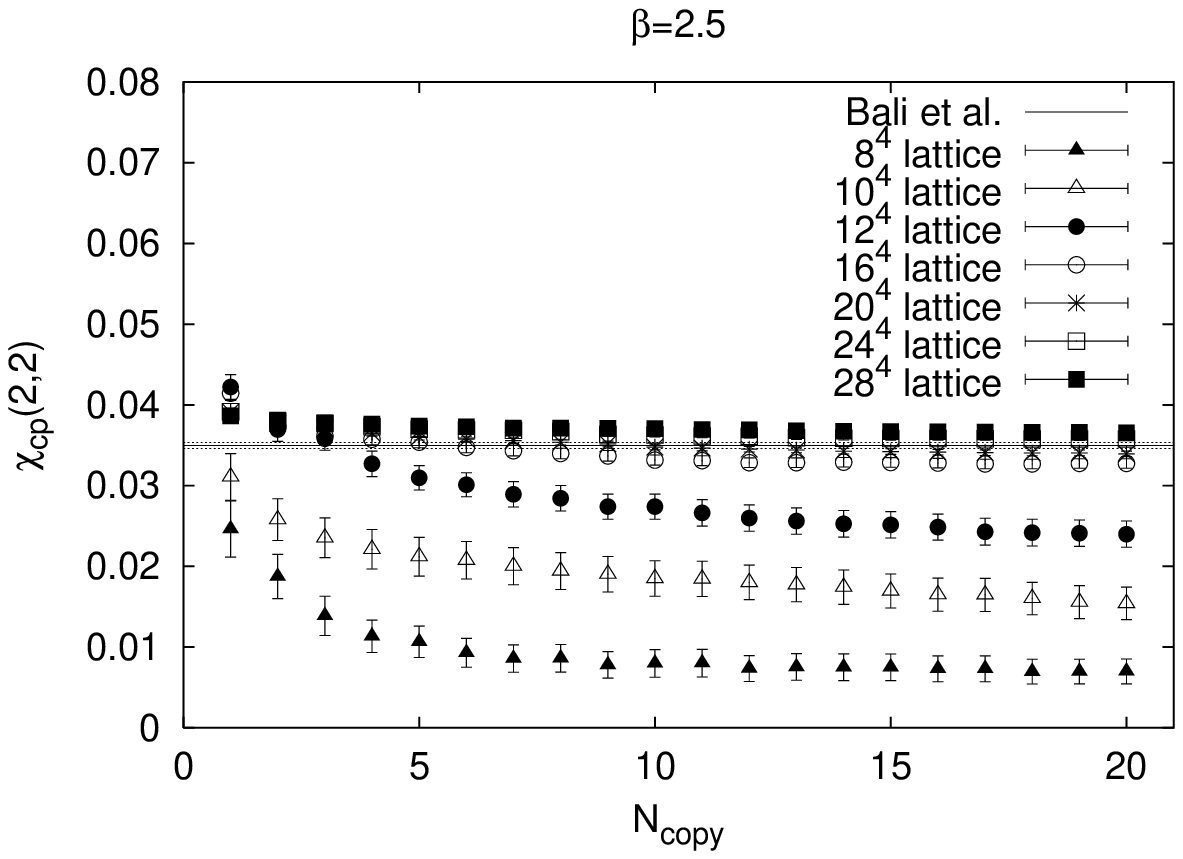}} &
\scalebox{0.6}{\includegraphics{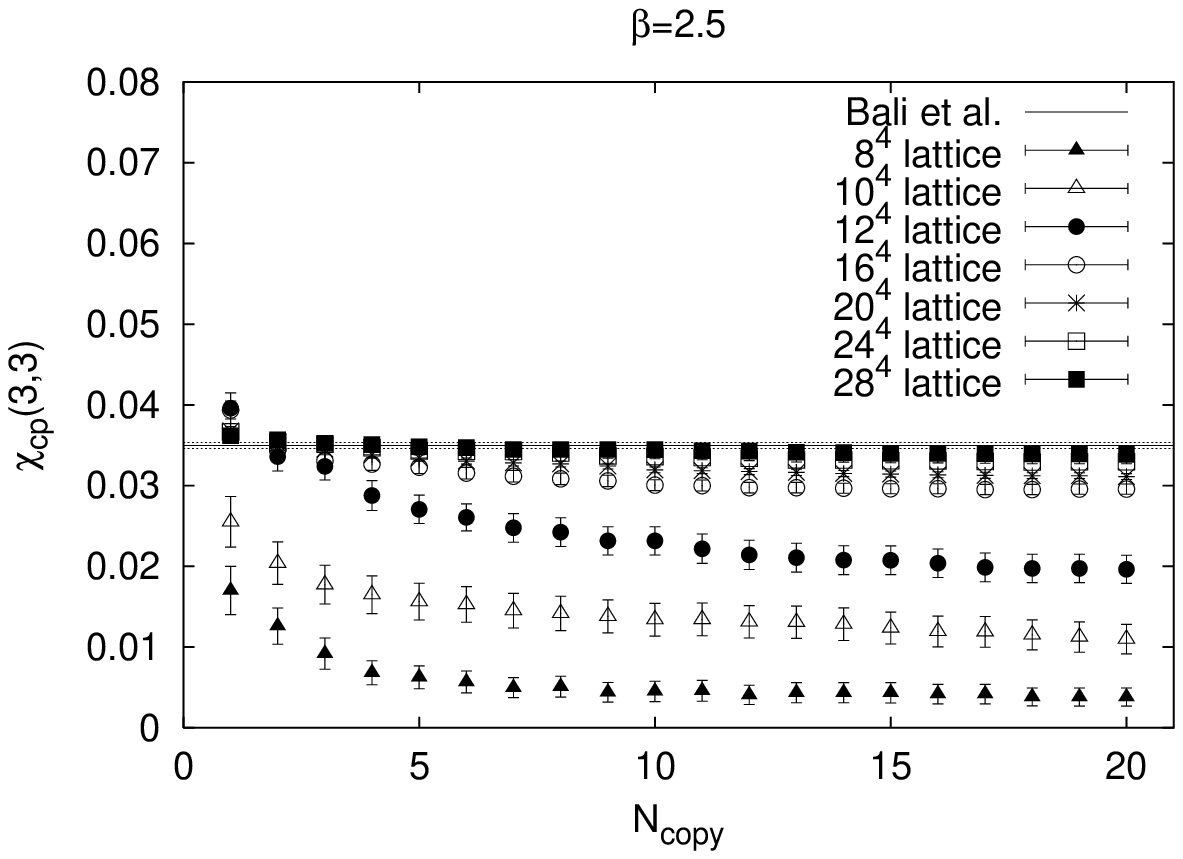}} \\
\scalebox{0.6}{\includegraphics{MCG_plt_b2p5c4.eps}} &
\scalebox{0.6}{\includegraphics{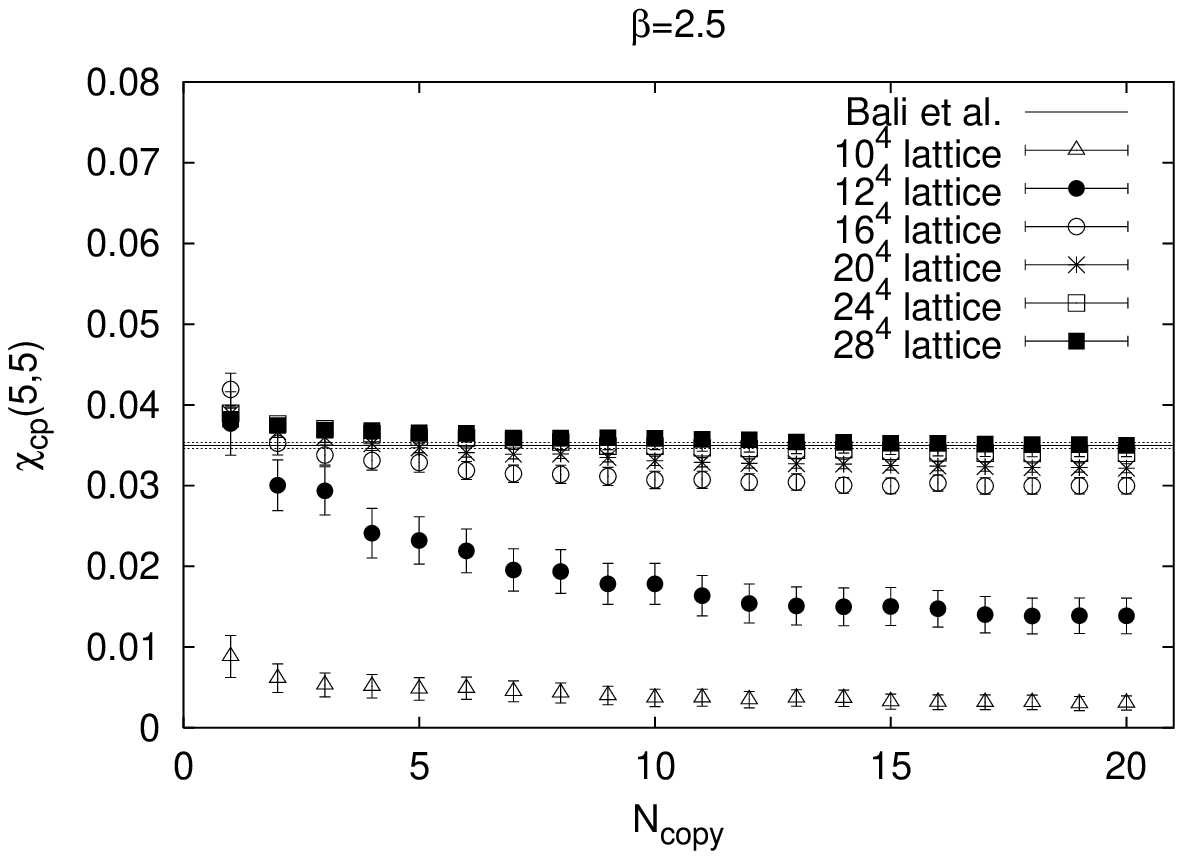}} \\
\end{tabular}
\caption{Center projected Creutz ratios $\chi_{cp}^{N_{copy}}(I,I)$ 
vs.\ $N_{copy}$ and lattice volume, at $\b=2.5$ and $I=2-5$.}
\label{more25}
}   

  In Fig.\ \ref{creutz} we show the projected Creutz ratios
$\chi_{cp}^{N_{copy}}(I,I)$ for $N_{copy}=5,10,15,20$ 
on the largest lattices we
have used: $20^4$ at $\b=2.3$ and $\b=2.4$, and $28^4$ at $\b=2.5$. We
also show the values of these Creutz ratios
extrapolated to the $N_{copy} \ra \infty$ limit, using the fitting
function \rf{fit} suggested by BKPV.  As usual in maximal center
gauge, all the $\chi_{cp}(I,I)$ for $I\ge 2$ are close to the
asymptotic string tension, and these latest results are not far from
our earlier results reported in ref.\ \cite{Jan98}.  As another way
of showing lattice size dependence, we take the average of the projected
($N_{copy}\ra \infty$) Creutz ratios $\chi_{cp}(I,I)$ in the range $I=2-5$
\beq
        \chi_{av} \equiv {1\over 4}\sum_{I=2}^5 
                            \chi_{cp}(I,I)
\eeq
Figure \ref{chi_av} shows the lattice size dependence of $\chi_{av}$.
Note the approach of $\chi_{av}$ to the full asymptotic string tension
($\s_{SU(2)}=0.035$ at $\b=2.5$) as lattice size
increases.

\FIGURE[h]{
\centerline{\scalebox{0.9}{\includegraphics{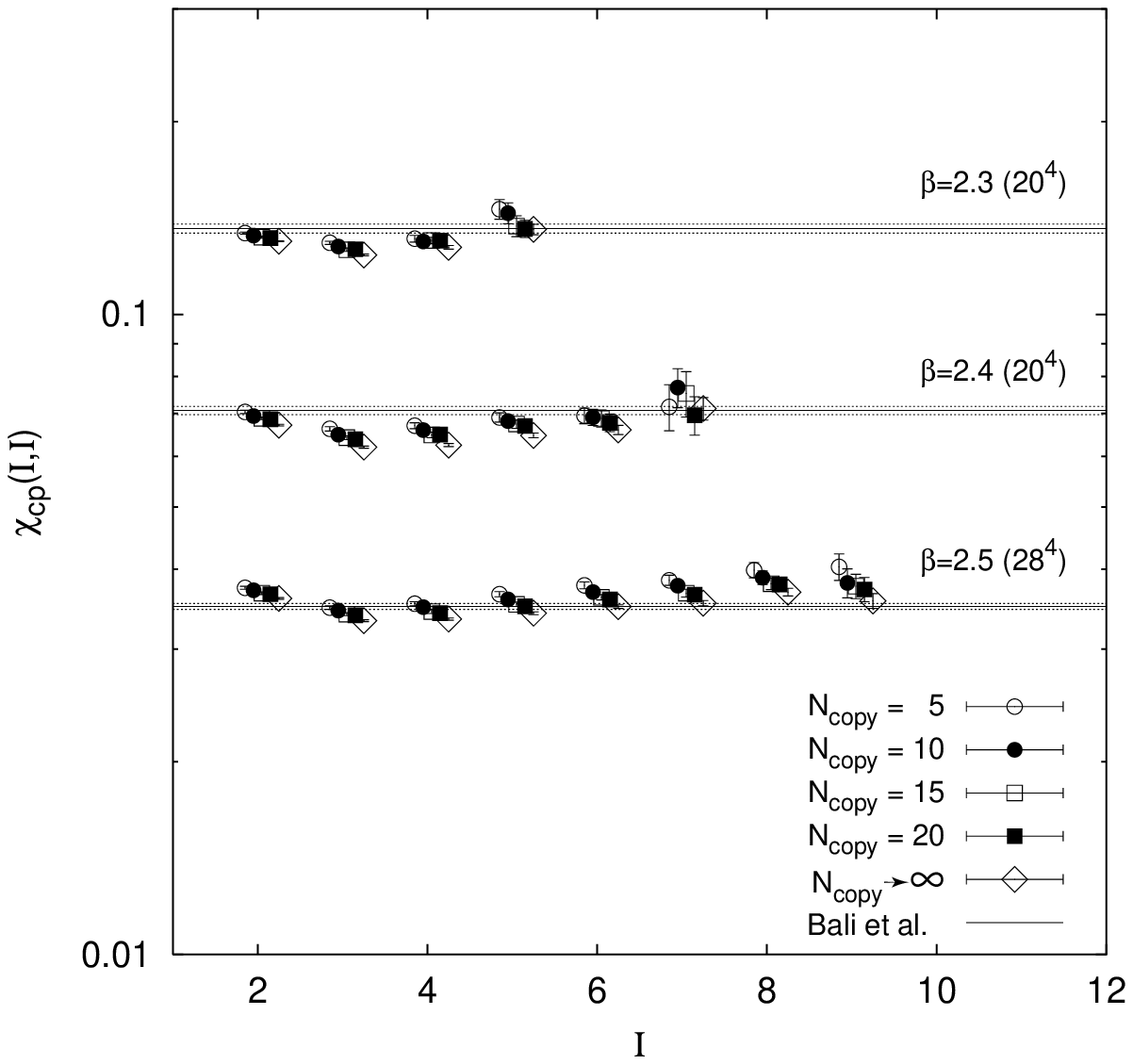}}}
\caption{Center-projected Creutz ratios $\chi_{cp}(I,I)$ at 
$N_{copy}=5,10,15,20$, and extrapolated to $N_{copy}\ra \infty$, on
the largest lattices ($20^4$ lattices at $\b=2.3,2.4$, and $28^4$ 
at $\b=2.5$) used in our simulations.}
\label{creutz}
}    

\FIGURE[h]{
\centerline{\scalebox{0.9}{\includegraphics{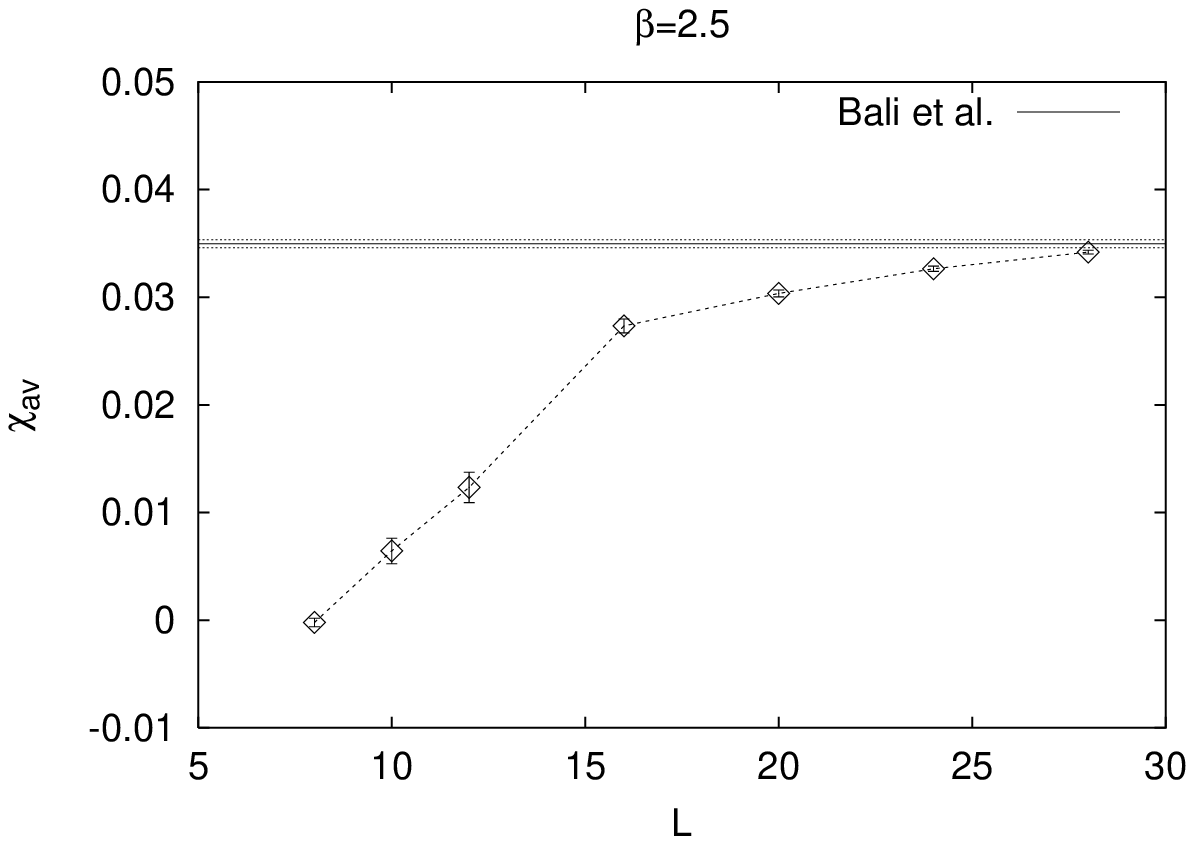}}}
\caption{Average of projected, $N_{copy}\ra \infty$ extrapolated
Creutz ratios $\chi_{cp}(I,I)$, in the range $I=2-5$.}
\label{chi_av}
}

\subsection{Gauge-Fixing Convergence Criterion}

   In addition to lattice volume and $N_{copy}$ dependence, it
is also worthwhile to check that the numerical results are stable
when the gauge-fixing convergence criterion is strengthened, i.e.\
when the constant $\d$ in eq.\ \rf{delta} is reduced.  When $\d$
is chosen too large, the center projected Creutz ratios come out
significantly too high.  This is illustrated in Fig.\ \ref{svh},
which shows results for projected Creutz ratios with convergence criteria
$\d=10^{-2},10^{-3},10^{-4},2\times 10^{-7}$ at $\b=2.5$ ($24^4$ lattice).  
The weakest convergence criterion,
corresponding to $\d=10^{-2}$, is clearly insufficient for
accurate results, but Creutz ratios
obtained with the two smallest values of $\d$ are fairly consistent,
indicating that these numbers are not far from the $\d \ra 0$ limit.

\FIGURE[h]{
\centerline{\scalebox{0.9}{\includegraphics{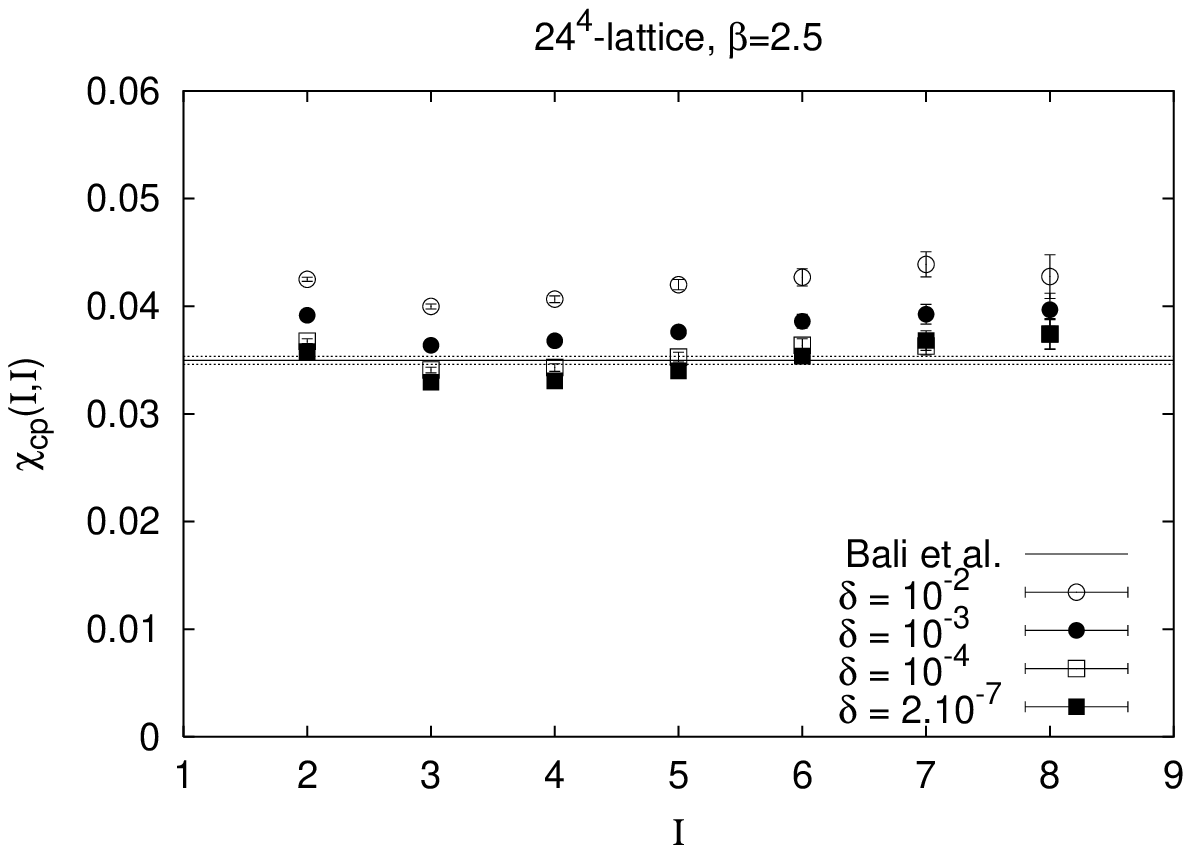}}}
\caption{Effect of varying the convergence criterion $\d$ on
projected Creutz ratios (extrapolated to $N_{copy}\ra \infty$)
at $\b=2.5$ on a $24^4$ lattice.}
\label{svh}
}

\section{Vortex Density}

  Next we consider the scaling of the vortex density.  The lattice P-vortex
density $p$ is the total number of P-vortex plaquettes (i.e.\ plaquettes
on the projected lattice with $ZZZZ = -1$), 
divided by the total number of plaquettes on the lattice.
This quantity is proportional to the average area taken up by P-vortices 
per unit lattice volume, and is determined from the center-projected plaquette
expectation value
\beq
       p = \oh (1 - W_{cp}[1,1])
\eeq
If this quantity scales as predicted by asymptotic freedom, then
we would have $p=p_{af}$, where
\beq
       p_{af} = 
   {\rho \over 6 \Lambda^2} F(\b)
\eeq
Here $\rho$ is the vortex density (average vortex area per
unit volume) in physical units of inverse area, and
\beq
      F(\b) =\left({6\pi^2 \over 11}\b\right)^{102/121}
               \exp\left[-{6\pi^2 \over 11}\b\right]
\label{F}
\eeq
In Fig.\ \ref{rho} we plot the lattice vortex density rescaled by the
asymptotic freedom expression
\beq
               \tilde{p} \equiv  {p \over F(\b)}
\eeq
which should be constant in the large $\b$ limit, if $p$ scales
according to asymptotic freedom.  There appears to be good evidence
for this kind of scaling, already for $\b \ge 2.2$, in agreement
with previous results \cite{Jan98,Tubby1}.  In Fig.\
\ref{rho} and Table \ref{tab1}, the vortex densities at $\b=2.3,2.4,2.5$ 
are taken from the largest lattices ($20^4$, $20^4$ and $28^4$, respectively),
and extrapolated to the $N_{copy}\ra \infty$ limit by a fit to
\beq
       p^{N_{copy}} = p + {c \over \sqrt{N_{copy}}}
\eeq
The vortex densities at other values of $\b$ (with $N_{copy}=3$) are
just taken from our previous work.  For comparison, in Table \ref{tab1},
we display our values
for $\tilde{p}$, and the values of the rescaled asymptotic string
tension
\beq
      \tilde{\s} \equiv \s_{SU(2)}/F(\b)
\eeq
where $\s_{SU(2)}$ is the string tension (in lattice units) on the
unprojected lattice, at the given $\b$ value.
It is interesting to note that the scaling
of vortex density $p$, in the range $\b = 2.3-2.5$, is substantially
better than the the scaling of the full asymptotic string tension
$\s_{SU(2)}$ in this range.

\TABLE[h!]{
\centerline{
\begin{tabular}{|c|c|c|} \hline
  $\b$    &  $\tilde{p}$   &  $\tilde{\s}$   \\ \hline 
  $2.3$   & $2.51(2)\times 10^3$ & $3.89(6)\times 10^3$ \\   
  $2.4$   & $2.47(3)\times 10^3$ & $3.34(5)\times 10^3$ \\   
  $2.5$   & $2.35(3)\times 10^3$ & $2.73(3)\times 10^3$ \\ \hline 
\end{tabular} } 
\caption{Rescaled vortex density $\tilde{p}$ and SU(2) string
tension $\tilde{\s}$.  These quantities should be constant
in the scaling limit.}
\label{tab1}
}


\FIGURE[h]{
\centerline{\scalebox{0.9}{\includegraphics{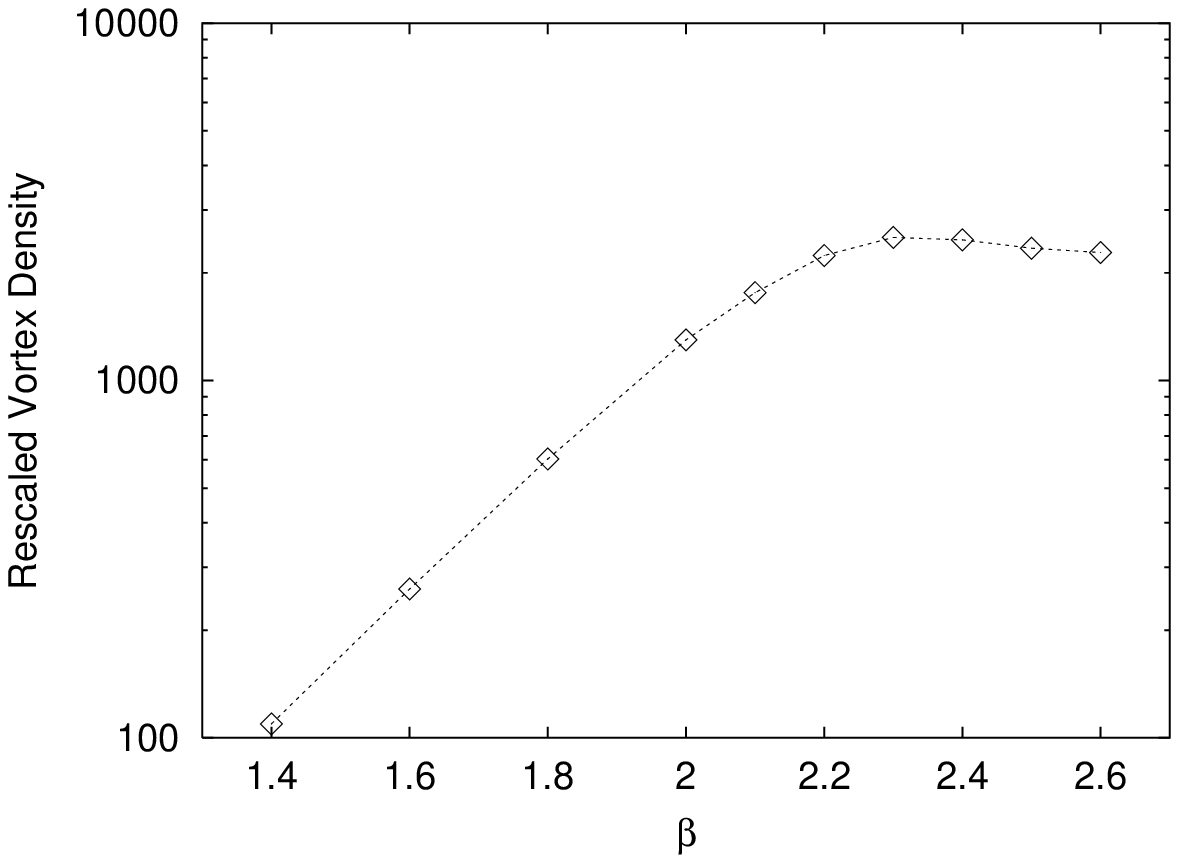}}}
\caption{Rescaled vortex density $\tilde{p}=p/F(\b )$, where 
$p$ is the measured vortex density and $F(\b)$ is the asymptotic
freedom expression in eq.\ \rf{F}.  Vortex density scales
if $\tilde{p}$ is constant.}
\label{rho}
}    

\section{Vortex Thickness}

   Why are such relatively large lattices required, in order to avoid
large finite size effects in center projected Creutz ratios and vortex
densities?  We believe that the relevant length scale here is
associated with the vortex thickness.  Center vortices are
surface-like objects, in $D=4$ dimensions, which have a finite
thickness in physical units.  The thick vortex surface (or ``core'')
bounds a Dirac 3-volume, which represents the region of discontinuity
of a singular gauge transformation associated with the vortex.  In
ref.\ \cite{vf} we have explained the vortex-finding property of
maximal center gauge in terms of the global properties of this
singular gauge transformation.  On a small lattice, however, with an
extension comparable to the vortex thickness, these global aspects of
the vortex field may be almost absent, and the minimal Dirac 3-volume
could be quite small.  In that case, the argument of ref.\ \cite{vf}
breaks down.  For this reason, we expect center projection to be less
effective at finding vortices on small lattices, leading to
underestimates of both the vortex density and the projected Creutz
ratios.

   Assuming there is some truth in this explanation, center
projection can only be accurate for lattices whose extension is large compared
to the vortex thickness.  There are three independent ways of estimating
the thickness of center vortices, which can be deduced from either
\begin{enumerate}
\item the ratio of ``vortex-limited'' Wilson loops \cite{Jan98};
\item the vortex free energy as a function of lattice size \cite{ET};
\item the adjoint string-breaking length \cite{dFP}.
\end{enumerate}
Vortex-limited Wilson loops are defined in the following way: $W_n(C)$
is a Wilson loop evaluated on a sub-ensemble of unprojected
configurations, selected 
so that precisely $n$ P-vortices, in the corresponding center-projected
configurations, pierce the minimal area of the loop.  We can further
make the restriction, for $W_1(C)$, that the negative P-vortex plaquette
lies at (or touches) the center point of the rectangular loop.  It is
then expected that 
\beq
      {W_1(C) \over W_0(C)} \ra -1
\eeq
in the limit where the vortex core is entirely contained within
the loop (cf.\ ref.\ \cite{Jan98} for a more extended discussion).  In
Fig.\ \ref{nvtex} we show the data for $W_1/W_0$ vs.\ loop area
at $\b=2.3$, taken from our previous work in ref.\ \cite{Jan98}.
Judging from this figure, the vortex appears to almost fit
inside a $5\times 5$ loop, which leads to a rough estimate of the
vortex radius, as it pierces a plane, of about 3 lattice spacings.
At $\b=2.3$ we have $\s a^2 = 0.135$, and taking $\s = 5~\mbox{fm}^{-2}$,
the lattice spacing is $a=0.164$ fm.  A diameter of 6 lattice
spacings then corresponds to a vortex thickness of about one fermi.

\FIGURE[h]{
\centerline{\scalebox{0.6}{\includegraphics{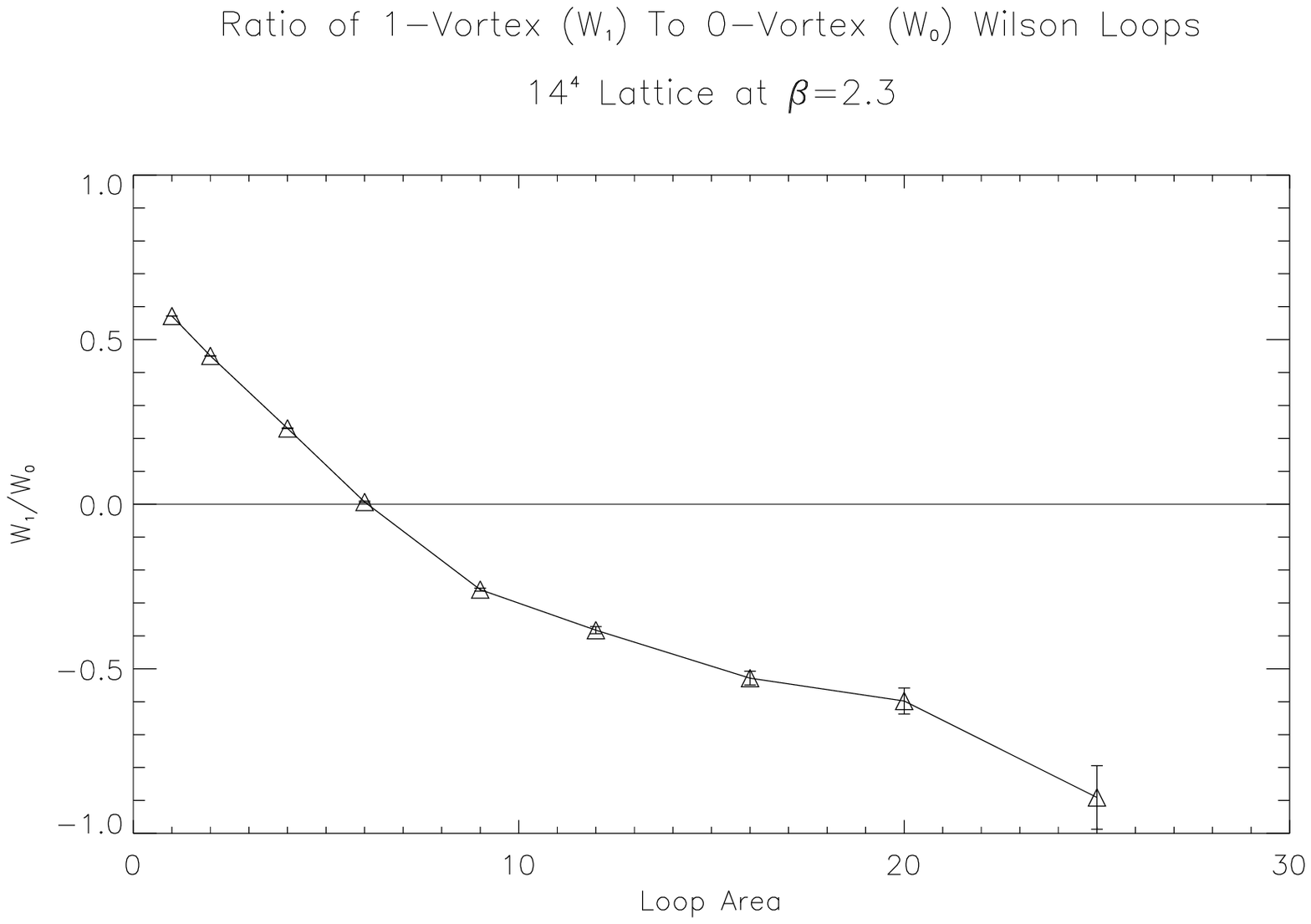}}}
\caption{Ratio of the 1-Vortex to the 0-Vortex Wilson loops
$W_1(C)/W_0(0)$, vs.\ loop area at $\b=2.3$.}
\label{nvtex}
}    

   A second estimate is obtained from the recent calculation of
vortex free energy vs.\ lattice size, carried out numerically 
by Kov\'{a}cs and Tomboulis \cite{ET}. The vortex free energy is 
close to zero when the lattice extension is greater than the vortex
thickness, and this again gives an estimate for the vortex thickness of
a little over one fermi.  Finally, if confinement is due to center
vortices, then an $R\times T$ Wilson loop in the
adjoint representation must change from a (Casimir scaling) area-law
falloff to a (color-screening) perimeter-law falloff for charge
separation $R$ greater than the vortex thickness \cite{Cas}.  
The adjoint string-breaking distance has been measured, by
de Forcrand and Philipsen \cite{dFE}, to be $1.25$ fm, and this distance
provides us with a third estimate of the vortex thickness, which is
roughly consistent with the other two.

   At $\b=2.5$, one fermi corresponds to 12 lattice spacings.  The lattice
used by BKPV at $\b=2.5$ was only 16 lattice spacings across, and this
may simply be inadequate for center projection to reliably identify
vortices, in view of the above 
estimates for the vortex thickness.  In fact, the P-vortex density is
quite low on small lattices, increasing sharply
up to $L=16$ lattice spacings at $\b=2.5$, where it begins
to level off.  This is illustrated in Fig.\ \ref{rxL}; the data points
are the extrapolated values for $p$ at $N_{copy}\ra \infty$.  
 
\FIGURE[h]{
\centerline{\scalebox{0.9}{\includegraphics{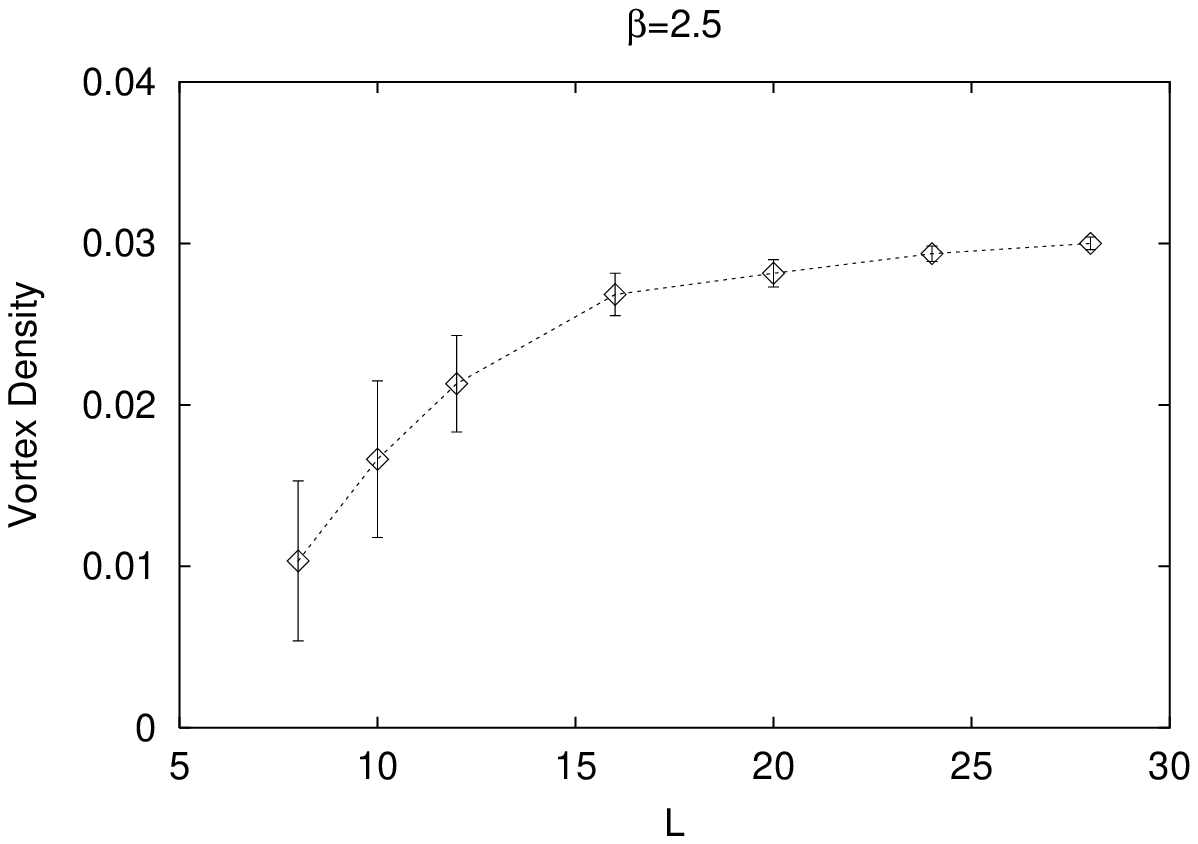}}}
\caption{Vortex density $p$ vs.\ the extension of $L$ of
the hypercubic lattice (volume $L^4$) at $\b=2.5$.}
\label{rxL}
}    

   In order to estimate the average distance between vortices, we
need to know the vortex density.  The P-vortex density, discussed in
the previous section, is in fact an overestimate of the
actual center vortex density \cite{Tubby1,bertle}. 
The reason is that P-vortices fluctuate within the
thick vortex core, and P-vortex locations, while correlated
among random Gribov copies, do vary somewhat from one random copy to 
another \cite{Jan98}. Thus, although P-vortices are certainly near the middle
of center vortices, they are unlikely to be exactly in the middle.
A more accurate estimate of the center vortex
density is arrived at by either ``smoothing'' the P-vortex surfaces, as
explained in ref.\ \cite{bertle}, or else by taking the naive estimate
of the number of vortices piercing a plane (per unit area
in lattice units), extracted
directly from the string tension (see, e.g., ref.\ \cite{Cas})
\beq
        f = {1\over 2}(1 - e^{-\s})
\eeq
The two estimates agree fairly well \cite{bertle}, and for $\b=2.3$ we
find $f=0.063$.  This implies an average distance of $f^{-1/2}\approx
4$ lattice spacings between the centers of vortex cores piercing a
plane.  Since we have already estimated the vortex thickness at
$\b=2.3$ to be about 6 lattice spacings, its clear that there must be
a substantial overlap between vortex cores, even on a very large
lattice.  There is nothing in principle wrong with that; vortex cores
are not impenetrable objects, and their long-range effects are
associated with Dirac
3-volumes, rather than the detailed structure
of the core.   Our findings for vortex thickness and
separation simply indicate, in accordance with some old ideas of the
Copenhagen group (Nielsen, Olesen, and Ambj{\o}rn in ref.\ \cite{tH}),
that the QCD vacuum is more like a liquid of vortices than a dilute
gas.  Perhaps this is natural for objects which are both condensed and
disordered.

\section{Conclusions}

   We have found that center-projected lattices are more sensitive to
finite size effects than are unprojected lattices, and precision
results for center-projected Creutz ratios require lattice sizes which
are large compared to one fermi.  If accurate numbers (rather than
just qualitative results) are required, then center-projected data
from numerical simulations must be tested for convergence with respect
to: (i) increasing the lattice size; (ii) increasing the number
$N_{copy}$ of gauge copies used for center gauge fixing, and (iii)
strengthening the gauge-fixing convergence criterion.

   Our work was stimulated by the recent findings of BKPV \cite{Borny},
who have demonstrated a significant $N_{copy}$-dependence for center-projected
Creutz ratios obtained on relatively small lattice volumes.  In the
large $N_{copy}$ limit, a large deviation ($\approx 30\%$) was found
between the string tension on unprojected and projected lattices.  Our
result in the present article is that this $N_{copy}$ dependence is greatly
reduced as lattice size increases, and center projected Creutz ratios
$\chi_{cp}(I,I)$ appear to converge to values which are quite close to
the asymptotic string tension obtained on unprojected lattices. 

   It is difficult to say whether the full and center-projected
string tensions are in precise agreement in the infinite volume
limit.  To check such agreement, what is really required is an
extrapolation to a triple limit: $\mbox{volume} \ra \infty$, $N_{copy} 
\ra \infty$, and $\d \ra 0$; our data is not yet adequate to
extrapolate to this triple limit systematically.
But we emphasize again that on the largest lattices we have used, 
and with results extrapolated
(following BKPV) to the $N_{copy}\ra \infty$ limit, the center projected
Creutz ratios lie quite close to the asymptotic string tension, as
illustrated in Fig.\ \ref{creutz} above.  Moreover, on the same large lattices
and $N_{copy}\ra \infty$ extrapolation, 
there is good evidence for asymptotic scaling of the vortex density at
couplings $\b > 2.2$.  The scaling of the vortex density, at the $\b$ values 
studied, is in fact substantially better than the scaling of string tension 
itself.

   Center projection in maximal center gauge is intended as a method
for locating center vortices, on unprojected lattices, from the position
of P-vortices on projected lattices.  The correspondence of P-vortices
with physical objects depends on
\begin{enumerate}
\item Scaling of the P-vortex density;
\item Correlation of P-vortex locations on the projected lattice 
with gauge-invariant observables on the unprojected lattice.
\end{enumerate}
In particular, it is the correlation of P-vortices with unprojected
Wilson loops, and specifically the ratios of ``vortex-limited'' Wilson
loops $W_n(C)/W_0(C)$ on the unprojected lattice, 
which indicate that P-vortices
correspond to center vortices (rather than to some other type of
object) on the unprojected lattice \cite{mog}.  Moreover, if
P-vortices locate center vortices, and if center vortices produce the
full asymptotic string tension, then we must find
\begin{enumerate}
\item[3.] Center dominance; i.e.\ the equality of the string tension
on projected and unprojected lattices.
\end{enumerate}
The numerical simulations reported here were concerned with the
scaling of the P-vortex density, and with center dominance.  From the
results of those simulations, we conclude that center projection in maximal
center gauge remains a viable method of locating center vortices on
full, unprojected lattices, and that these vortices probably account
for the entire asymptotic string tension.

\acknowledgments{ Our research is supported in part by Fonds zur
F\"orderung der Wissenschaftlichen Forschung P13997-PHY (M.F.), the
U.S. Department of Energy under Grant No.\ DE-FG03-92ER40711 (J.G.),
the ``Action Austria-Slovakia: Cooperation in Science and Education''
(Project No.\ 30s12) and the Slovak Grant Agency for Science, Grant
No. 2/7119/2000 (\v{S}.O.). J.G. also acknowledges the support of
MaPhySto, Centre for Mathematical Physics and Stochastics, funded by
the Danish National Research Foundation.}

\end{document}